\documentclass[twocolumn,amsmath,amssymb]{snp}
\usepackage{epsfig}
\usepackage{hyperref}
\usepackage{url}
\usepackage{graphicx}
\usepackage{dcolumn}
\usepackage{bm}
\usepackage{graphicx,amssymb}
\usepackage{color}
\usepackage{float}
\newcommand{\pt}{$p_{ {\mathrm T} }$}
\newcommand{\y}{$y$}

\usepackage{multirow}
\usepackage{rotating}
\usepackage{lineno}
\usepackage{subfigure}
\begin{document}

\title{\Large Inclusive J/$\psi$ and $\psi$(2S) production in pp collisions at $\sqrt{s} = 7$ TeV at forward rapidity with ALICE at LHC}

\author{\large Biswarup Paul (For the ALICE Collaboration)}
\email{biswarup.paul@cern.ch}
\affiliation{Saha Institute of Nuclear Physics, Kolkata - 700064, India}
\maketitle

The ALICE detector is described in detail in~\cite{R1}.~The Muon Spectrometer of ALICE is designed to measure the $\psi$ and $\Upsilon$ states (J/$\psi$,$\psi(2S)$ and $\Upsilon(1S)$,$\Upsilon(2S)$,$\Upsilon(3S)$) in the forward pseudo-rapidity interval of $-$4 $\leq \eta \leq$ $-$2.5.

The present analysis uses pp collisions data at $\sqrt{s} = 7$ TeV. The data were recorded in 2011 with a trigger defined by the coincidence of a minimum bias trigger with the detection of two opposite sign muons reconstructed in the trigger chambers of the muon spectrometer. A total of 4 million events were analysed, corresponding to an integrated luminosity $\mathcal{L}_{\rm int}$ = 1.35 pb$^{-1}$ (with 5\% systematic uncertainty). In order to improve the purity of the muon tracks the following selection criteria were applied: 
(1) both muon tracks match with trigger tracks above the 1 GeV/$\it{c}$ \pt\ threshold,
(2) both muon tracks in the pseudo-rapidity range $-$4 $\leq \eta \leq$ $-$2.5,
(3) transverse radius coordinate of the track at the end of the absorber (longitudinal position of absorber from interation point (IP) is $-$5.0 $\leq$ $z$ $\leq$ $-$0.9 m) in the range 17.6 $\leq$ $R_{\rm{abs}}$ $\leq$ 89.5 cm.
(4) dimuon rapidity in the range  2.5 $\leq$ \y\ $\leq$ 4.0.
(5) dimuon \pt\ in the range  0 $\leq$ \pt\ $\leq$ 20 GeV/$\it{c}$.

\begin{figure}[h]
\vspace{-4.00mm}
\includegraphics[scale=0.32]{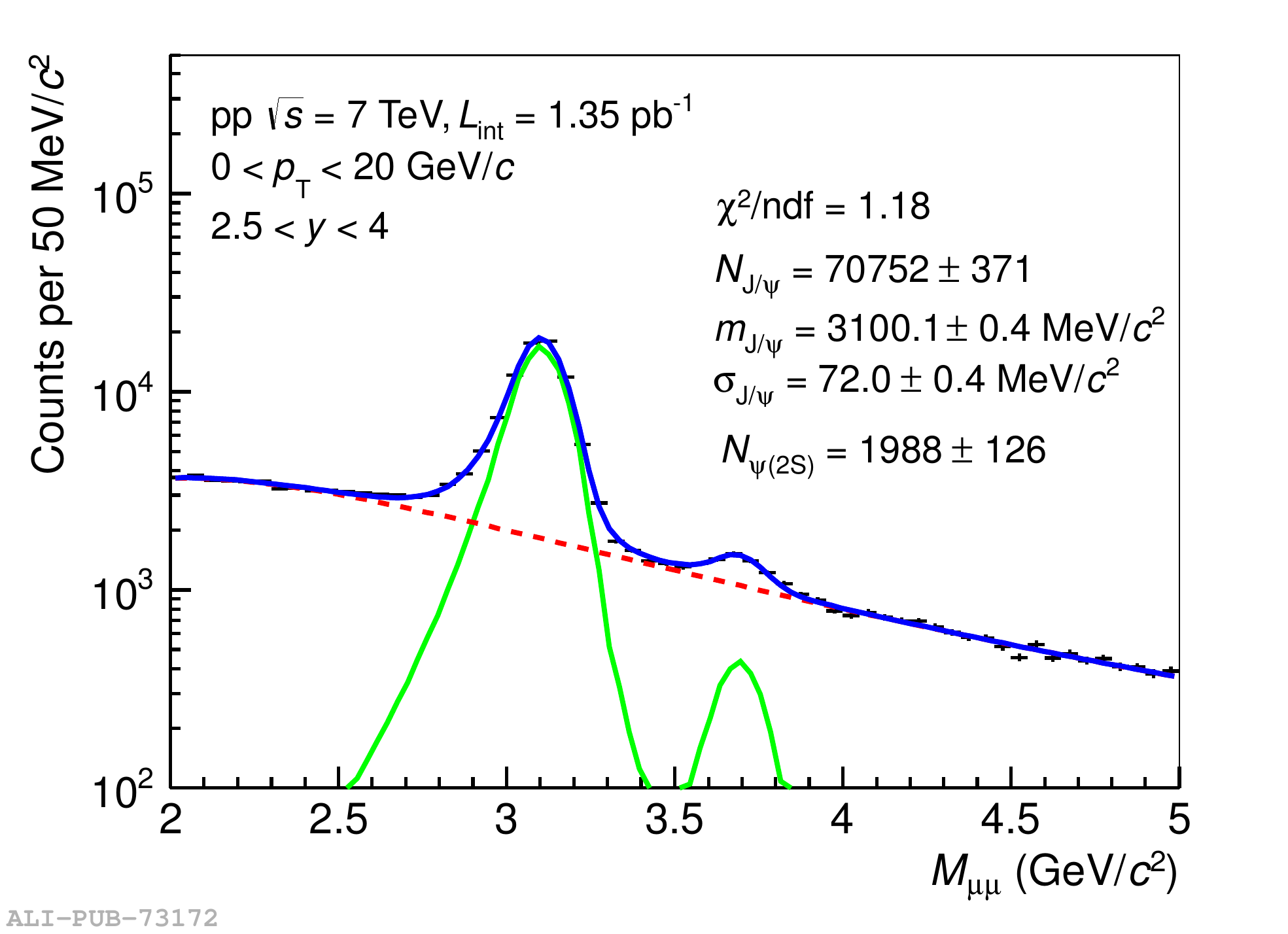}
\vspace{-4.00mm}
\caption{\label{fig1} Opposite sign dimuon invariant mass spectra, integrated over \y\ (2.5 $\leq$ \y\ $\leq$ 4.0) and \pt\ (0 $\leq$ \pt\ $\leq$ 20 GeV/$\it{c}$)~\cite{R5}.}
\end{figure}

Fig.~\ref{fig1} shows the invariant mass spectra fitted with two Extended Crystal Ball functions (Crystal Ball function with a non gaussian tail on both sides) for two signals and a variable width gaussian function (a gaussian function with a width varying linearly with the mass) for the background.
\begin{figure}[h]
\vspace{-4.00mm}
\includegraphics[scale=0.33]{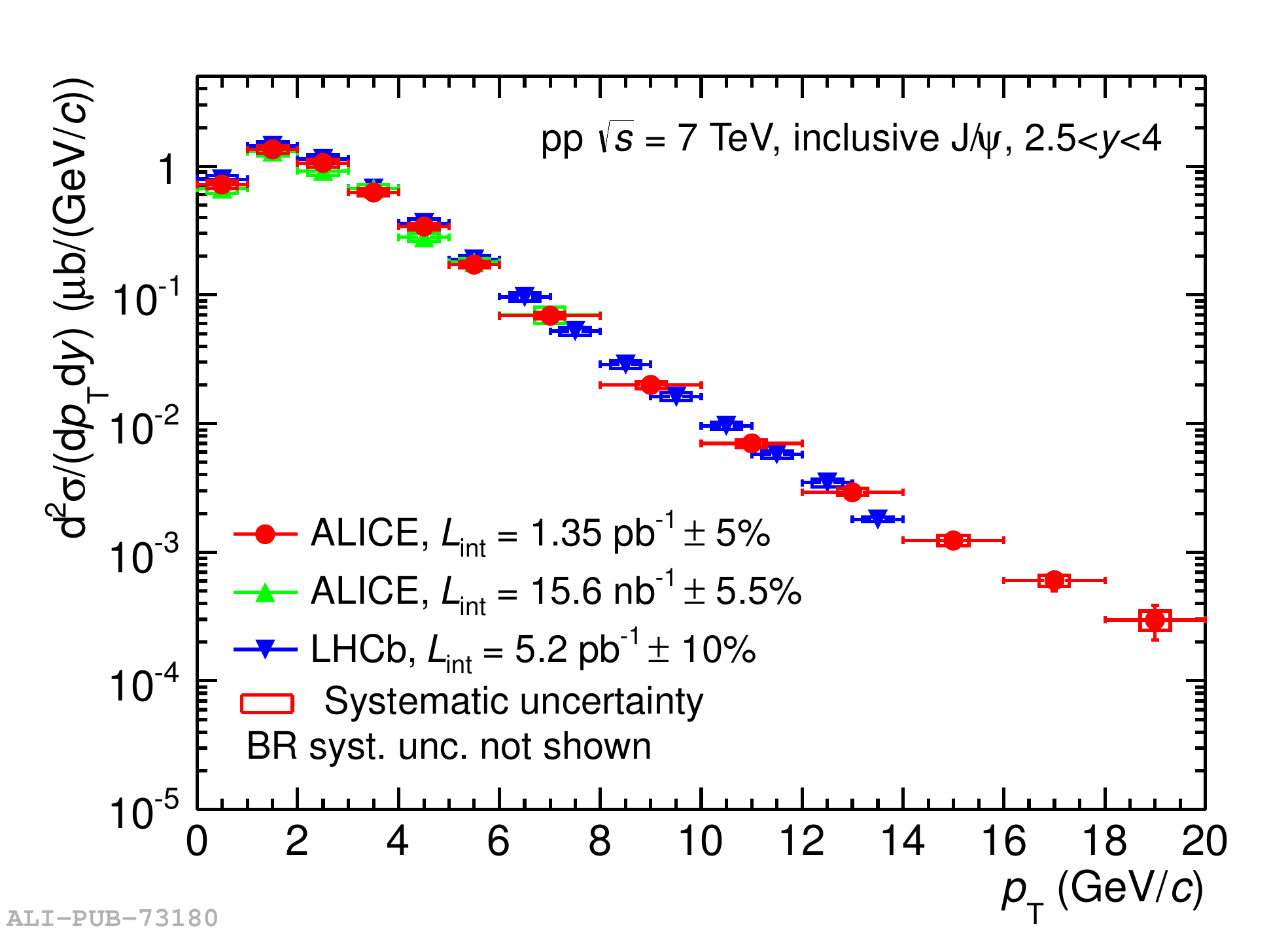}
\vspace{-4.00mm}
\caption{\label{fig2} \pt\ differential cross section of J/$\psi$~\cite{R5}.}
\end{figure}
\begin{figure}[!]
\vspace{-4.00mm}
\includegraphics[scale=0.33]{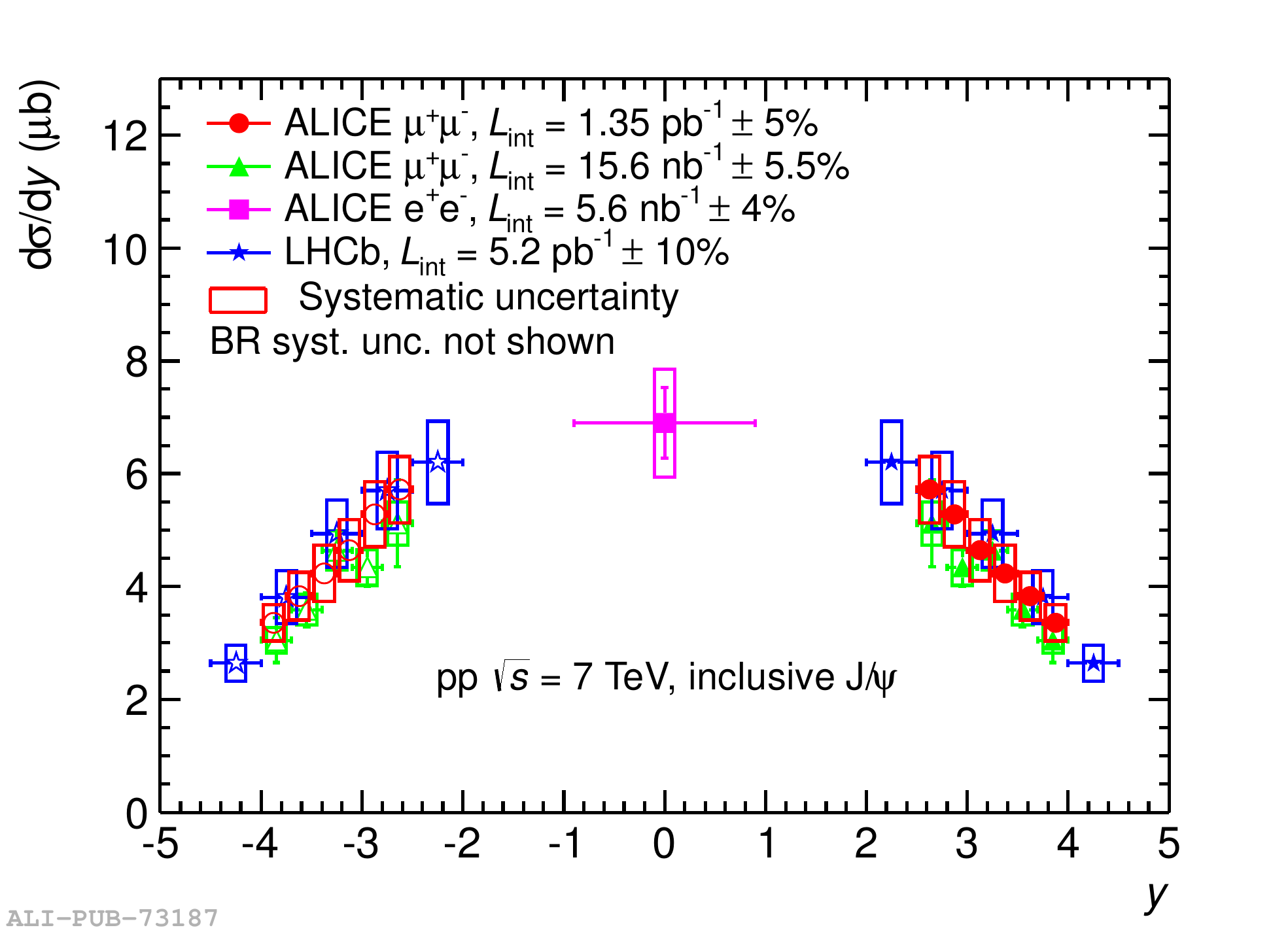}
\vspace{-4.00mm}
\caption{\label{fig3} \y\ differential cross section of J/$\psi$~\cite{R5}.}
\end{figure}

We present here inclusive production of J/$\psi$ and $\psi$(2S). Inclusive measurements contain, on top of the direct production, contributions from the decay of higher excited states as well as contributions from non-prompt production. The production cross sections of J/$\psi$ and $\psi$(2S) were determined by normalizing the production yields (the measured yields from the fits to the invariant mass spectra were corrected by the acceptance times efficiency factor) with the branching ratio and the integrated luminosity. The systematic uncertainties on the cross sections arise due to signal extraction, MC parametrization, trigger and tracking efficiency, matching efficiency and luminosity determinations.

The measured production cross sections of J/$\psi$ and $\psi$(2S), integrated in the \y\ and \pt\ range are:\\
$\sigma_{\rm J/\psi}$~=~6.69~$\pm$~0.04~(stat.)~$\pm$~0.63~(syst.)~$\mu$b.\\
$\sigma_{\rm \psi(2S)}$~=~1.13~$\pm$~0.07~(stat.)~$\pm$~0.19~(syst.)~$\mu$b.
\vspace{-3.00mm}
\begin{figure}
\vspace{-4.00mm}
\includegraphics[scale=0.33]{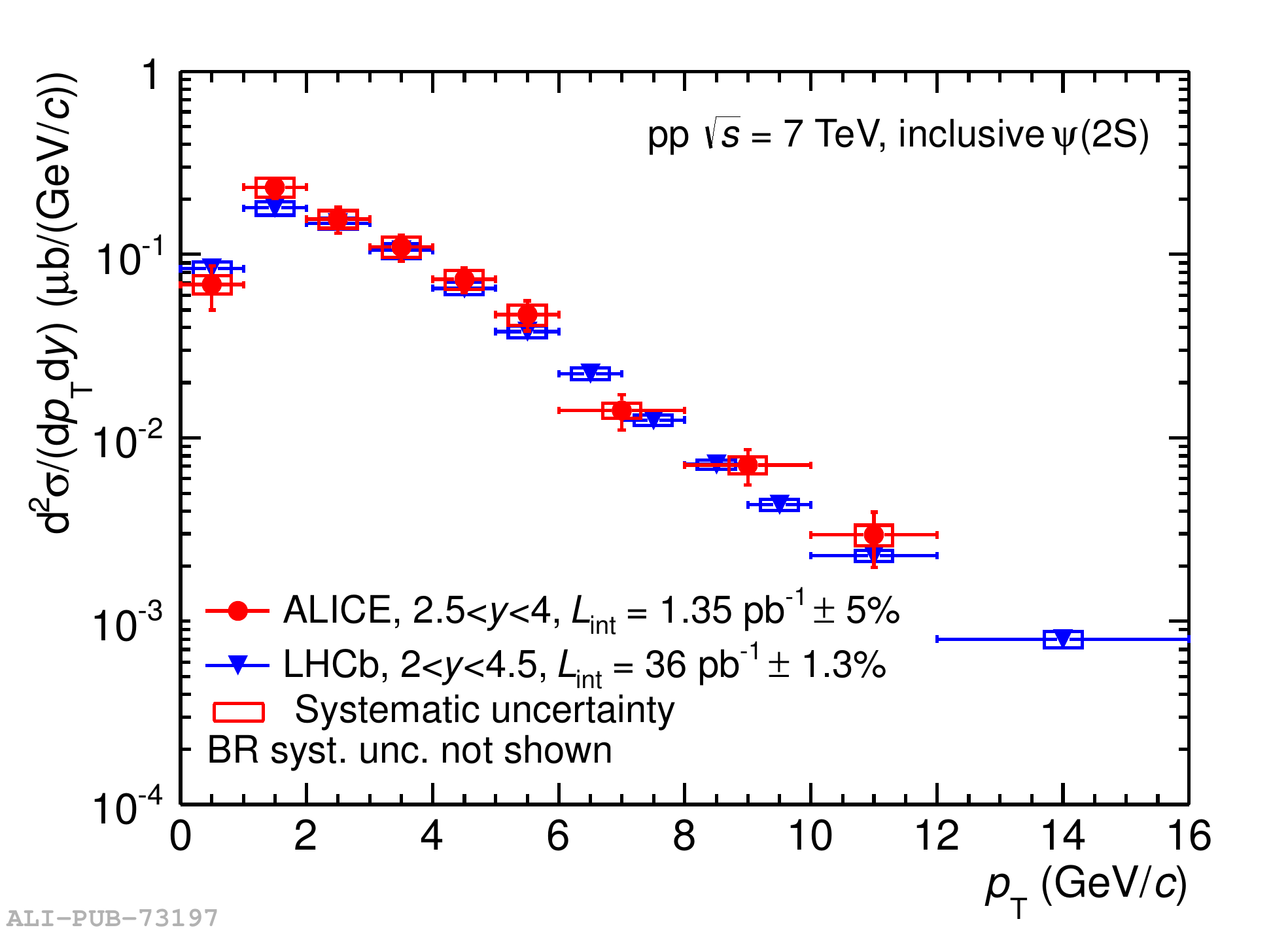}
\vspace{-4.00mm}
\caption{\label{fig4} \pt\ differential cross section of $\psi$(2S)~\cite{R5}.}
\end{figure}
\begin{figure}
\vspace{-4.00mm}
\includegraphics[scale=0.33]{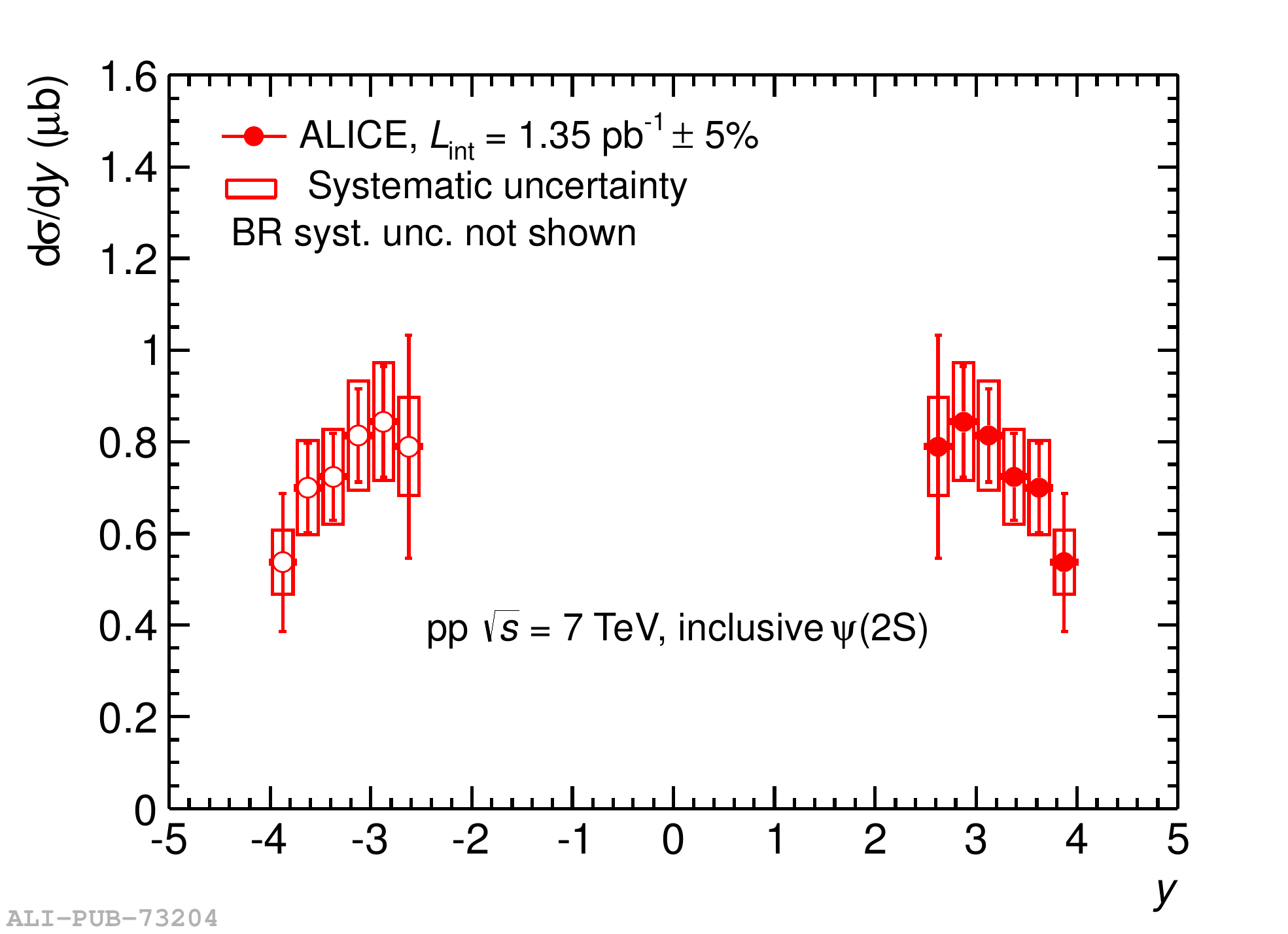}
\vspace{-4.00mm}
\caption{\label{fig5} \y\ differential cross section of $\psi$(2S)~\cite{R5}.}
\end{figure}

Fig.~\ref{fig2} and Fig.~\ref{fig3} show the differential production cross section of J/$\psi$ in thirteen \pt\ bins and in six \y\ bins, respectively. This result is consistent with the previous ALICE result~\cite{R2} and also with the measurement performed by the LHCb collaboration~\cite{R3}. This measurement extends J/$\psi$ cross section to 20 GeV/$\it{c}$ in \pt\ at forward rapidity.
\begin{figure}
\vspace{-4.00mm}
\includegraphics[scale=0.33]{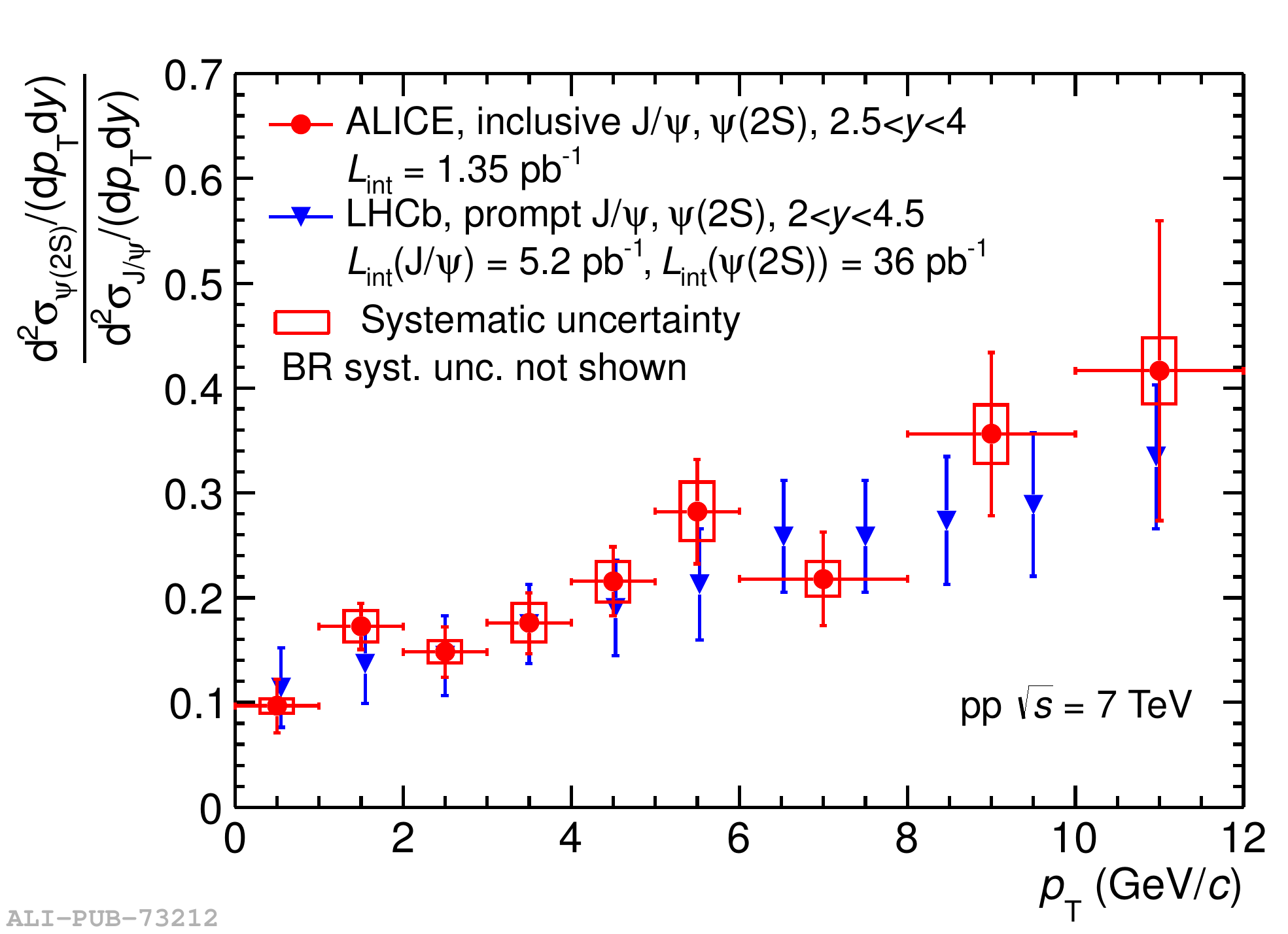}
\vspace{-4.00mm}
\caption{\label{fig6} $\psi$(2S)/J/$\psi$ ratio as a function of \pt~\cite{R5}.}
\end{figure}
\begin{figure}
\vspace{-4.00mm}
\includegraphics[scale=0.33]{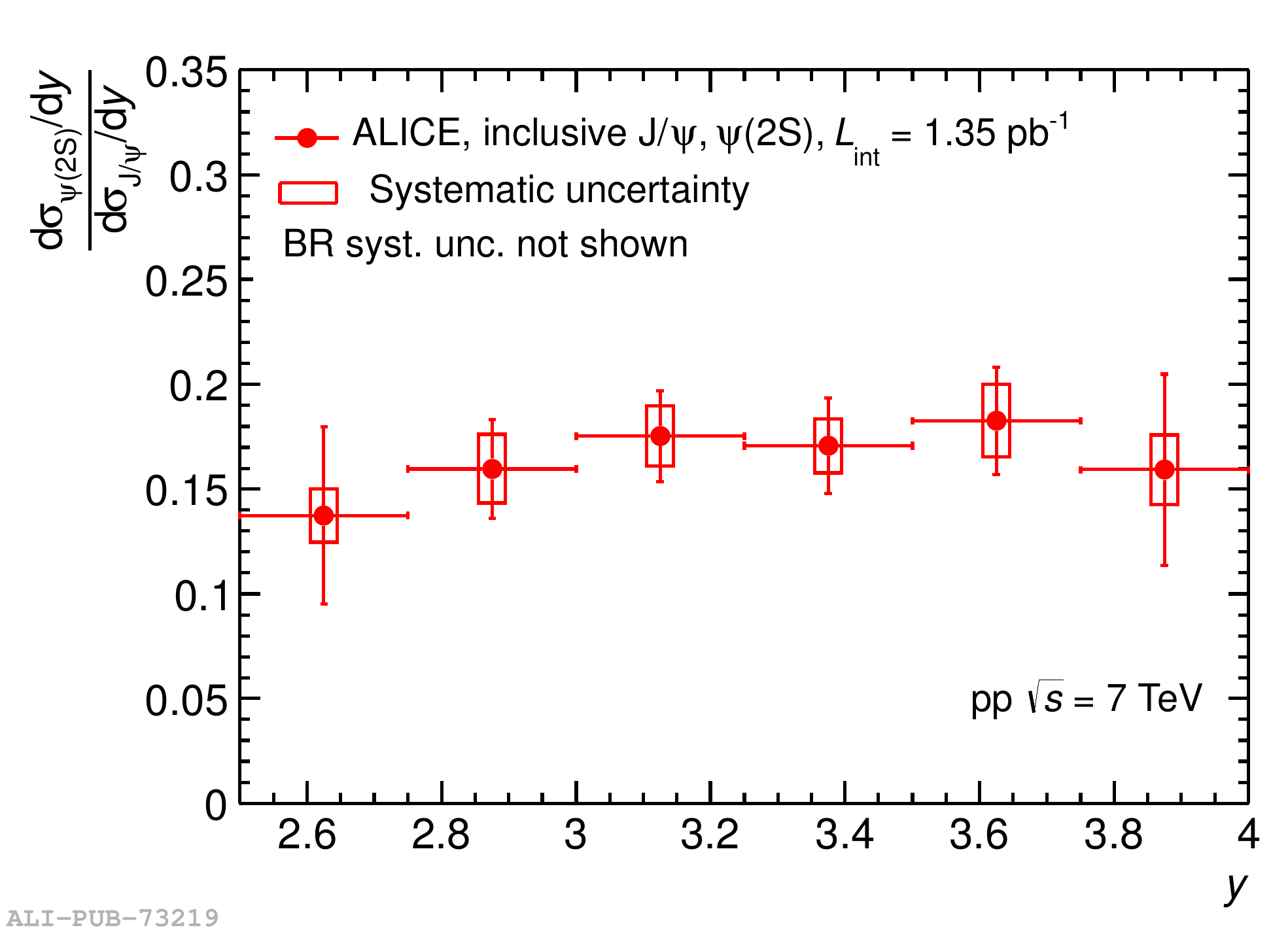}
\vspace{-4.00mm}
\caption{\label{fig7} $\psi$(2S)/J/$\psi$ ratio as a function of \y~\cite{R5}.}
\end{figure}

Fig.~\ref{fig4} shows the differential production cross section of $\psi$(2S) in nine \pt\ bins. The result is consistent with LHCb measurement~\cite{R4} in the same rapidity interval. Fig.~\ref{fig5} shows the differential production cross sections of $\psi$(2S) in six \y\ bins. This is the first $\psi$(2S) measurement in pp collisions at ALICE. 

The inclusive $\psi$(2S)/J/$\psi$ ratio, integrated over \pt\ and \y\ is 0.170 $\pm$ 0.011 (stat.) $\pm$ 0.013 (syst.). The $\psi$(2S)/J/$\psi$ ratio were measured as a function of \pt\ and \y\ as shown in Fig.~\ref{fig6} and Fig.~\ref{fig7} and a clear \pt\ dependence can be observed, in consistent with LHCb~\cite{R4}. No strong \y\ dependence is visible, in the \y\ range covered by the ALICE muon spectrometer. More details of this analysis can be found in~\cite{R5}.

\end{document}